\documentclass[prl,twocolumn,nofootinbib,showpacs,superscriptaddress]{revtex4-1}

\usepackage{amsfonts}
\usepackage{amsmath}
\usepackage{amssymb}
\usepackage{bm}
\usepackage{dcolumn}
\usepackage{epsfig}
\usepackage{graphicx}
\usepackage{graphics}
\usepackage[latin1]{inputenc}
\usepackage{latexsym}
\usepackage{rotating}
\usepackage[colorlinks=true]{hyperref}
\usepackage[all]{hypcap}
\usepackage{xspace} 
\usepackage[usenames]{color}

\usepackage{ulem}
\definecolor {darkgreen}{rgb}{0.2,0.7,0.2}

\newcommand\be{\begin{equation}}
\newcommand\ba{\begin{eqnarray}}
\newcommand\ee{\end{equation}}
\newcommand\ea{\end{eqnarray}}

\newcommand\bw{\begin{widetext}}
\newcommand\ew{\end{widetext}}

\newcommand{\nn}{\nonumber}

\newcommand{\GR}{{\mbox{\tiny GR}}}

\newcommand{\MAT}{{\mbox{\tiny mat}}}

\newcommand{\PPE}{{\mbox{\tiny PPE}}}
\newcommand{\E}{{\mbox{\tiny E}}}
\newcommand{\Edot}{{\mbox{\tiny $\mathcal{F}$}}}
\newcommand{\U}{{\mbox{\tiny U}}}
\newcommand{\K}{{\mbox{\tiny K}}}

\newcommand{\DD}{{\mbox{\tiny DD}}}

\newcommand{\mrm}{\mathrm}

\begin{document}
\title{Gravitational Waves from Quasi-Circular Black Hole Binaries \\ in Dynamical Chern-Simons Gravity}

\author{Kent Yagi}
\affiliation{Department of Physics, Montana State University, Bozeman, MT 59717, USA.}

\author{Nicol\'as Yunes}
\affiliation{Department of Physics, Montana State University, Bozeman, MT 59717, USA.}

\author{Takahiro Tanaka}
\affiliation{Yukawa Institute for Theoretical Physics, Kyoto University, Kyoto, 606-8502, Japan.}

\date{\today}

\begin{abstract} 

Dynamical Chern-Simons gravity cannot be strongly constrained with current experiments because it reduces to General Relativity in the weak-field limit. This theory, however, introduces modifications in the non-linear, dynamical regime, and thus, it could be greatly constrained with gravitational waves from the late inspiral of black hole binaries. We complete the first self-consistent calculation of such gravitational waves in this theory. For favorable spin-orientations, advanced ground-based detectors may improve existing solar-system constraints by 6 orders of magnitude.
 
\end{abstract}



\maketitle

{\emph{Introduction}}.--- General Relativity (GR) has been tested to exquisite accuracy in the Solar System and with binary pulsars~\cite{will-living}, constraining any possible deviations when the gravitational field is relatively weak and mildly dynamical. The non-linear, dynamical, strong-field regime of GR, where the quadrupole formula is insufficient to describe the dynamics, remains mostly observationally unconstrained. Gravitational waves (GWs) emitted during the late inspiral and merger of compact objects would be ideal probes of this regime. The second-generation of ground-based GW detectors (Adv.~LIGO~\cite{ligo}, Adv.~Virgo~\cite{virgo}, KAGRA~\cite{kagra} and LIGO-India~\cite{Schutz:2011tw}) will allow for the first strong-field tests of GR.

Modified gravity theories that reduce exactly to GR in the weak-field, yet deviate in the strong-field, exist. One example is dynamical Chern-Simons (CS) gravity~\cite{jackiw,CSreview}, where the Einstein-Hilbert action is modified by the product of a dynamical scalar field and the Pontryagin topological invariant. Such a correction is unavoidable in superstring theory~\cite{alexander:2004:lfg,Alexander:2004xd,polchinski2} and it also arises naturally in loop quantum gravity~\cite{taveras,Mercuri:2009zt}, and in effective field theories of inflation~\cite{weinberg-CS}.

Dynamical CS gravity has not been heavily constrained by experiments because (i) it only interacts with gravity and (ii) it reduces exactly to GR in the weak-field limit. This theory has the same post-Newtonian parameters as GR in the weak-field~\cite{Alexander:2007zg,Alexander:2007vt}. The leading-order non-vanishing modification to the motion of bodies enters through corrections to Lense-Thirring precession~\cite{alihaimoud-chen}. LAGEOS~\cite{LAGEOS} and Gravity Probe B~\cite{GPB} can thus constrain the theory, but only extremely weakly.   

GWs emitted during the late inspiral and coalescence of black hole (BH) binaries encode invaluable information about the fundamental gravitational interaction in the strong field, precisely where CS deviations are largest. Supermassive  BH mergers are not ideal for such tests because their radius of curvature is large, and thus, CS corrections are naturally suppressed~\cite{sopuertayunes,panietal,prisgair}. On the other hand, stellar-mass BHs, the targets of ground-based GW detectors, are ideal for testing CS gravity because their radius of curvature is small.

Future GW tests of dynamical CS gravity require the calculation of waveform templates with which to filter GW data~\cite{lrr-2005-3}. This is essential because the signal-to-noise ratio (SNR) of ground-based detectors is expected to be low, at least initially. Therefore, the calculation of templates in dynamical CS gravity is urgent, if we are to test this theory with future GW observations. In calculating such templates, we will discover how a BH binary shrinks due to the loss of energy to GW and scalar radiation, how the binary's binding energy is modified due to the presence of a scalar field and how Kepler's third law is corrected. These results are relevant to other astrophysical tests of dynamical CS gravity, for example with low-mass x-ray binaries~\cite{kent-LMXB}.

{\emph{Dynamical Chern-Simons Gravity}}.--- This theory is defined by the action~\cite{CSreview}
\be
\!S \! = \!\!\! \int \!\! d^4x \sqrt{-g} \left(\!\! \kappa_g R \!+\! \frac{\alpha}{4} \vartheta R_{\nu\mu \rho \sigma} \!{}^* R^{\mu\nu\rho\sigma} \!\!\! -\! \frac{\beta}{2} \nabla_\mu \vartheta \nabla^{\mu} \vartheta \!\!+\!\! \mathcal{L}_{\MAT} \right)\!,
\label{action}
\ee
where $\kappa_g \equiv (16\pi G)^{-1}$, $g$ is the determinant of the metric $g_{\mu\nu}$, $R_{\mu\nu \delta \sigma}$ and ${}^* R^{\mu\nu\rho\sigma}$ are the Riemann tensor and its dual, $R$ is the Ricci scalar, $\vartheta$ is a dynamical field, $(\alpha,\beta)$ are coupling constants and $\mathcal{L}_{\MAT}$ is the matter Lagrangian density. We define the dimensionless parameter $\zeta \equiv \xi/m^4$, where $\xi \equiv \alpha^2/(\kappa_g \beta)$ and $m$ is the total mass of the system. The characteristic length scale of the theory is given by $\xi^{1/4}$ and Solar System tests require $\xi^{1/4} \leq \mathcal{O}(10^8)$ km~\cite{alihaimoud-chen} (see~\cite{CSreview} for more details).

{\emph{Adiabatic Quasi-Circular BH Inspirals}}.--- The inspiral of comparable-mass compact objects can be studied within post-Newtonian (PN) theory, where one assumes all characteristic velocities are much smaller than the speed of light and gravitational fields are weak~\cite{blanchet-review}. We here concentrate on quasi-circular orbits, because, by the time GWs emitted in generic orbits enter the sensitivity band of ground-based detectors, they will have circularized due to GW emission~\cite{Peters:1963ux,Peters:1964zz}. 

A circular orbit is fully described by its binding energy $E$. In dynamical CS gravity, this quantity contains three contributions: a gravitational potential energy $E_{\U}$, a kinetic energy $E_{\K}$ and a scalar interaction energy $E_{\DD}$. $E_{\U}$ can be calculated, to leading PN order, via $\int U'_1 \rho'_2 d^3x'$, where the primes mean that $U_{1}$ and $\rho_{2}$ are functions of $x'^i$, $U_{A}$ is the gravitational potential of BH $A$~\cite{kent-CSBH}
\be
U_A \equiv - \frac{m_A}{r_A} \left( 1 + 3 \frac{Q_A^{ij}}{m_A} \frac{n_{A\left\langle ij \right\rangle}}{r_A^2} \right)\,, 
\label{pot}
\ee
and $\rho_{A}$ is the density of BH $A$
\be
\rho_A \equiv \left( m_A + Q_A^{ij} \partial_i \partial_j \right) \delta^{(3)}({x}^{k}-{x}_A^{k} )\,,
\label{rho}
\ee
with the quadrupole moment given by $Q_A^{ij}\equiv (201/3584) \zeta (m^4/m_{A}) \chi_A^2 \hat{S}_A^{\left\langle i \right.} \hat{S}_A^{\left. j \right\rangle}$~\cite{kent-CSBH}. Here, $m_{A}$ is the individual mass, $\chi_{A} = |S^{i}_{A}|/m_{A}$ is the dimensionless Kerr spin parameter, $\hat{S}_A^{i}$ is the spin angular momentum unit vector, and $r_{A}$ and $n^{i}_{A}$ are the field point distance and unit vector, all relative to the $A$th BH, with the angle-brackets representing the symmetric and trace-free operation, i.e.~$n_{A\left\langle ij \right\rangle} \equiv n_{Ai} n_{Aj} - (1/3) \delta_{ij} n_{Ak} n_{A}^k$. The potential in Eq.~\eqref{pot} can be read directly from the $(t,t)$ component of the metric of an isolated BH in dynamical CS gravity~\cite{kent-CSBH}. The density in Eq.~\eqref{rho} must be calculated by solving $\square U_{A} = 4 \pi \rho_{A}$, with the potential of Eq.~\eqref{pot}. Combining all these results,
\begin{align}
E_\U &= -\frac{\mu m}{2 r_{12}} \left\{ 1 - \frac{201}{1792} \zeta  \frac{m^2}{m_1^2} \chi_1^2 \left[ 1-3(\bm{n}_{12} \cdot \hat{\bm{S}}_1)^2 \right] \frac{m^{2}}{r_{12}^{2}} \right\}
\nn \\
& + (1 \leftrightarrow 2)\,,
\label{EU}
\end{align}
where $r_{12}$ and $n_{12}^{i}$ are the binary's orbital separation and the separation's unit vector, $\mu = m_{1} m_{2}/m$ is the reduced mass, and $(\bm{A} \cdot \bm{B})$ is the flat-space scalar inner product.

The scalar field has a rest energy and an interaction energy, which is induced because spinning BHs in dynamical CS gravity possess a magnetic-type dipole scalar field. When two such BHs are present, the dipole-dipole interaction energy is 
\be
E_\DD = -\frac{25}{256} \zeta \frac{m^4}{r_{12}^3} \chi_1 \chi_2 \left[ (\hat{\bm{S}}_1 \cdot \hat{\bm{S}}_2) - 3(\bm{n}_{12} \cdot \hat{\bm{S}}_1)(\bm{n}_{12} \cdot \hat{\bm{S}}_2) \right]\,.
\label{EDD}
\ee
This expression is derived by integrating out the interaction Lagrangian, namely the kinetic and effective source term for the scalar field. This result is derived by analogy with neutron stars with magnetic dipole moments in GR~\cite{ioka}, modulo a sign difference due to the difference between considering a vector field and a pseudo-scalar field. The kinetic energy to leading PN order is $E_\K = \mu v^2/2$, with $v$ the relative velocity. 

Let us now re-express $E_\U$, $E_\DD$ and $E_\K$ in terms of $u \equiv (\pi m f)^{1/3} = (m \omega)^{1/3}$ where $f$ is the GW frequency and $\omega$ is the orbital angular velocity. This is achieved by finding the leading-order, effective-one-body, equation of motion for the binary constituents, which in the center of mass frame is $\mu \, r_{12} \, \omega^2 = F_\U + F_\DD$.  The magnitude of the gravitational and dipole-dipole forces are computed by differentiating the potential and dipole-dipole energies with respect to $r_{12}$. From this, we obtain Kepler's third law $r_{12} = (m/u^2) (1 + \delta C_r u^4)$ with
\begin{align}
\label{deltaCr}
\delta C_r & \equiv  \frac{25}{512} \zeta \frac{\chi_1 \chi_2}{\eta}  \left[ (\hat{\bm{S}}_1 \cdot \hat{\bm{S}}_2)  - 3\left\langle (\bm{n}_{12} \cdot \hat{\bm{S}}_1)(\bm{n}_{12} \cdot \hat{\bm{S}}_2) \right\rangle_\omega \right] 
\nn \\
&- \frac{201}{3584} \zeta \frac{m^2}{m_1^2} \chi_1^2 \left[ 1-3\left\langle (\bm{n}_{12} \cdot \hat{\bm{S}}_1)^2 \right\rangle_{\omega} \right] + (1 \leftrightarrow 2)\,,
\end{align}
where we have orbit averaged, as implied by $\left\langle \cdots \right\rangle_\omega$.
We also obtain $v=r_{12} \omega = u(1+\delta C_r u^4)$.

We now have all the ingredients to compute the binding energy in the center of mass frame in terms of $u$. Combining $E_\U$, $E_\DD$ and $E_\K$ and taking the orbital average, we find $E=-(\mu/2) u^2 (1+\delta C_E u^4)$, with
\begin{align}
\delta C_{\E} & \equiv -\frac{25}{256} \zeta \frac{\chi_1 \chi_2}{\eta} \left[ (\hat{\bm{S}}_1 \cdot \hat{\bm{S}}_2)  - 3 \left\langle (\bm{n}_{12} \cdot \hat{\bm{S}}_1)(\bm{n}_{12} \cdot \hat{\bm{S}}_2) \right\rangle_\omega \right]
 \nn \\
& + \frac{201}{1792} \zeta  \frac{m^2}{m_1^2} \chi_1^2 \left[ 1-3\left\langle (\bm{n}_{12} \cdot \hat{\bm{S}}_1)^2 \right\rangle_{\omega} \right] + (1 \leftrightarrow 2)\,,
\end{align}
where $\eta = \mu/m$ is the symmetric mass ratio. This reduces to Eq.~(100) of~\cite{kent-CSBH} in the test particle limit.

For a circular orbit, the {\emph{balance law}} states that the rate of change of the binding energy $\dot{E}$ must be exactly balanced by the flux of scalar and gravitational radiation taken out to infinity and into any horizons $\mathcal{F}$. One can then show that $\dot{E} = -(32/5) \mu^2 v^4 \omega^2 [1 + \delta C_{\Edot} u^4] = -(32/5) \eta^2 u^{10} [1 + (\delta C_{\Edot} + 4 \delta C_r) u^4]$ where~\cite{quadratic} 
\begin{align}
\label{deltaCEdot}
\delta C_{\Edot} \equiv \frac{25}{24576} \zeta \frac{1}{\eta^2} \left[ \Delta^2 +27 \left\langle (\bm{\Delta} \cdot \hat{\bm{v}}_{12})^2 \right\rangle_\omega  \right]\,. 
\end{align}
Here, $\hat{v}_{12}^{i} = \hat{v}_{1}^{i} - \hat{v}_{2}^{i}$ is a unit vector pointing in the direction of the difference of the orbital velocities, $\hat{v}_{12}^{ij} \equiv \hat{v}_{12}^{i}\hat{v}_{12}^{j}$, and $\Delta^i \equiv (m_2/m) \chi_1 \hat{S}_1^i - (m_1/m) \chi_2 \hat{S}_2^i$. The first term is a CS correction due to the emission of scalar radiation, while the second one is a CS correction to the emission of gravitational radiation.

{\emph{GWs from Quasi-Circular BH Inspirals}}.--- The balance law allows us to write an evolution equation for the GW frequency: $\dot{f} = \dot{f}_{\GR} ( 1+ \delta C  u^4 )$ and to leading order in the PN approximation $\dot{f}_{\GR} \equiv (96/5) \pi^{8/3} \mathcal{M}^{5/3} f^{11/3} + \mathcal{O}(u^{13})$~\cite{blanchet-review} while $\delta C \equiv  \delta C_{\Edot} - 3 \delta C_{\E} + 4 \delta C_r$, or
\allowdisplaybreaks[4]
\begin{align}
\label{deltaC}
\delta C &= \frac{101555}{344064} \zeta \frac{m^2}{m_1^2} \chi_1^2 \left[ 1 - \frac{58833}{20311} \left( \hat{\bm{S}}_1 \cdot \hat{\bm{L}} \right)^2 \right] \nn \\
&  - \frac{12725}{49152} \zeta \frac{\chi_1 \chi_2}{\eta} \left[ \left( \hat{\bm{S}}_1 \cdot \hat{\bm{S}}_2 \right) \right. \nn \\
& \left. - \frac{1467}{509} \left( \hat{\bm{S}}_1 \cdot \hat{\bm{L}} \right) \left( \hat{\bm{S}}_2 \cdot \hat{\bm{L}} \right) \right] + (1 \leftrightarrow 2)\,.
\end{align}
We will here perform the orbital averaging implied in Eqs.~\eqref{deltaCr}--\eqref{deltaCEdot}, assuming that the precession timescale is much longer than the orbital timescale.
This evolution equation then gives the time-domain representation of the GW phase via $\phi (t) = \int 2 \pi (f/\dot{f})df$.

In the extraction of GWs from noisy data, it is customary to employ the Fourier transform of the GW response function. This quantity can be computed analytically via the stationary phase approximation~\cite{Bender,cutlerflanagan,Droz:1999qx,Yunes:2009yz}. Neglecting PN amplitude corrections to the GW response, the sky-averaged Fourier transform is $\tilde{h}(f) = {\cal{A}} f^{-7/6} \exp[i \Psi(f)]$. The overall amplitude is the usual GR quantity: ${\cal{A}} = 30^{-1/2} \pi^{-2/3} {\cal{M}}^{5/6} D_{L}^{-1}$, where ${\cal{M}} = \eta^{3/5} m$ is the chirp mass and $D_{L}$ is the luminosity distance. The Fourier phase $\Psi(f) = \Psi_{\GR}(f) + \delta \Psi(f)$, where $\Psi_{\GR}(f)$ is the GR result~\cite{arun35PN,arunbuonanno,blanchet3PN}, while 
\be
\delta \Psi(f) =\frac{3}{128} (\pi \mathcal{M} f)^{-5/3} \left( -10 \; \delta C  \; u^{4} \right)
\label{phase}
\ee
is the CS correction. This correction enters at 2PN order, and thus, it is degenerate with the spin-spin PN correction to the GR Fourier phase. Such a modified gravitational waveform is naturally contained in the parameterized post-Einsteinian (ppE) framework~\cite{PPE} of the form $\Psi_{\PPE} = \Psi_{\GR} + \beta_{\PPE} (\pi \mathcal{M} f)^{b_{\PPE}}$ as
\be
\beta_{\PPE} = - \frac{15}{64} \; \delta C \; \eta^{-4/5}, \quad  b_{\PPE} = -\frac{1}{3}\,. 
\ee
%

{\emph{Validity of Approximations}}.--- We have here employed several approximations. First, we assumed a {\emph{small coupling}}, i.e.~$\xi \ll M^{4}$ where $M$ is the smallest mass scale of the problem. This approximation is required because Eq.~\eqref{action} represents an {\emph{effective theory}}; an expansion to quadratic order in the curvature. The field equations derived from Eq.~\eqref{action} are only valid to linear order in $\zeta'$. Second, an exact, closed-form solution that represents a spinning BH in dynamical CS gravity is currently known only to ${\cal{O}}(\chi_{A}^{2})$ in a $\chi_{A} \ll 1$ expansion. Therefore, the potential energy in Eq.~\eqref{action} is formally only valid to ${\cal{O}}(\chi_{A}^{2})$, and thus, so is the waveform computed above.

Since scalar radiation is caused by scalar dipole charges, we can estimate the accuracy of the slow-rotation expansion by computing the dipole charge to all orders in $\chi$ and then comparing this to an expression truncated to ${\cal{O}}(\chi^{2})$.  The dipole charge $\mu_{A}$ is defined as the asymptotic coefficient of the $r^{-2} \, \cos{\theta}$ term in the solution for $\vartheta$, when considering an isolated spinning BH in dynamical CS gravity. The variation of the action with respect to $\vartheta$~\cite{teukolsky,yunespretorius} leads to the $\vartheta$ equation of motion, as given in Eqs.~$(10)$ of~\cite{kent-CSBH}. Employing a multipolar decomposition, we can solve this equation at dipole ($\ell=1$) harmonic order using Green's function methods and exactly obtain the scalar dipole charge as
\be
\mu_A^{\mathrm{(full)}} = \frac{\alpha}{\beta}\frac{2+2\chi_A^4-2\sqrt{1-\chi_A^2}-\chi_A^2(3-2\sqrt{1-\chi_A^2})}{2 \chi_A^3}\,.
\label{mufull}
\ee
For $|\chi_A| < 0.8$, the difference between $\mu_A^{\mathrm{(full)}}$ and its truncated expansion at ${\cal{O}}(\chi^{2})$ is always less than 10$\%$.

{\emph{Future Constraints with GW Observations}}.---Let us assume that a GW observation has been made and found consistent with GR. One can then ask how large $\zeta$ can be to be consistent with such an observation, thus placing a constraint on $\xi^{1/4}$ by performing a Fisher analysis~\cite{cutlerflanagan}. For sufficiently high SNR, the accuracy to which a given parameter $\theta^{a}$ can be measured can be estimated via $(\Delta \theta^{a}) = \sqrt{\left( \Gamma^{-1} \right)^{aa}}$, where 
\be
\Gamma_{ab} \equiv 4 \mrm{Re} \int^{f_\mrm{max}}_{f_\mrm{min}} \frac{\partial_a \tilde{h}(f) \partial_b \tilde{h}(f)}{S_n(f)} df\,,
\ee
is the Fisher matrix, partial derivatives are with respect to $\theta^{a}$, and $S_n(f)$ is the noise spectral density for Adv.~LIGO with the zero-detuned configuration~\cite{Ajith:2011ec}\footnote{In the published version of this article~\cite{Yagi:2012vf}, we used the wide-band configuration of Adv.~LIGO~\cite{cornishsampson}. However, the zero-detuned configuration is more appropriate for tests of GR that modify the waveform at high PN order, as shown in~\cite{Sampson:2014qqa,Chatziioannou:2015uea}, which is why it is employed here.} (also for Adv.~Virgo and KAGRA), ET~\cite{mishra}, LISA~\cite{bertibuonanno} and DECIGO/BBO~\cite{yagi:brane}. The limits of integration $f_\mrm{min}=\max (f_\mrm{low}, f_\mrm{1yr})$ and $f_\mrm{max}=\min (f_\mrm{high}, f_{\mathrm{end}})$, where $f_\mrm{low}$ and $f_\mrm{high}$ are lower and higher cutoff frequencies of a given detector, respectively, while $f_\mrm{1yr}$ is the GW frequency 1 year prior to coalescence and $f_{\mathrm{end}}$ is the frequency at the innermost stable circular orbit which can be obtained by solving $\hat{C}_0=0$ where $\hat{C}_0$ is given in Eq.~(1.5) of~\cite{favata}. 

Let us first concentrate on the proposed constraints using second-generation ground-based detectors such as Adv.~LIGO. Unfortunately, if the spins of the binaries are (anti-)aligned, these detectors do not seem to be sensitive enough to measure $\zeta$, within the small coupling approximation, due to degeneracies between $\zeta$ and $\chi_{A}$. If the binaries are precessing, however, certain degeneracies between the CS corrections and the spin-spin GR couplings would be broken. Since there is a CS correction in the spin-spin interaction, the precession equation is modified at 0.5PN order relative to the leading GR spin-orbit interaction. The effect of spin enters at 1.5PN order relative to the leading Newtonian term in the phase, and hence the CS correction to the phase due to precession appears at 2PN order, which is the same order as the correction in Eq.~\eqref{deltaC}. Therefore, if one were to perform a Fisher analysis for precessing binaries, one would be forced to first derive the CS modified precession equations and solve them numerically, together with the CS modified frequency evolution equation to obtain the time-domain waveform. One then needs to carry out a discrete Fourier transform with appropriate time-domain filters to avoid spectral leakage, and carefully take the parameters derivatives of the waveform numerically. 
Since the goal of this letter is to derive the CS corrected waveform, and with this, to provide a rough (order of magnitude) estimate of how well future detectors can constrain the theory, carrying out the above mentioned analysis is beyond the scope of this letter.

\begin{figure}[htb]
\begin{center}
\begin{tabular}{l}
\includegraphics[width=4.3cm,clip=true]{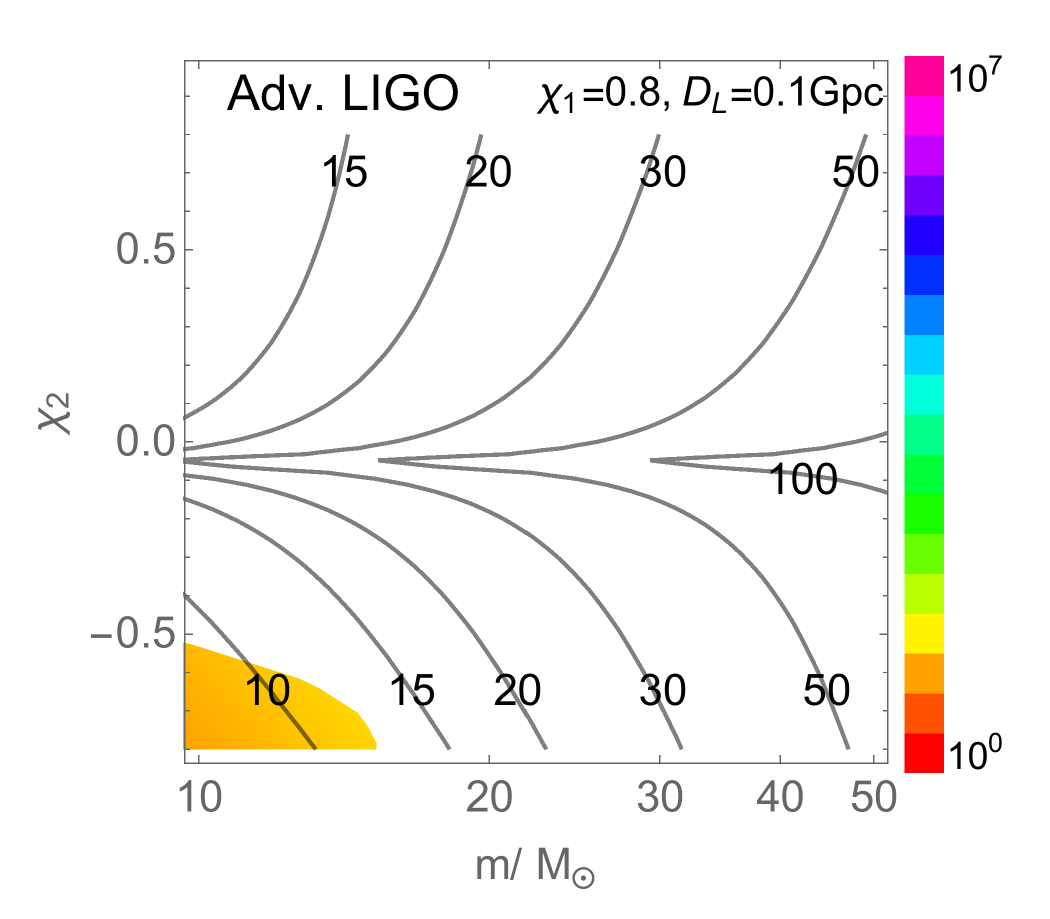}  
\includegraphics[width=4.3cm,clip=true]{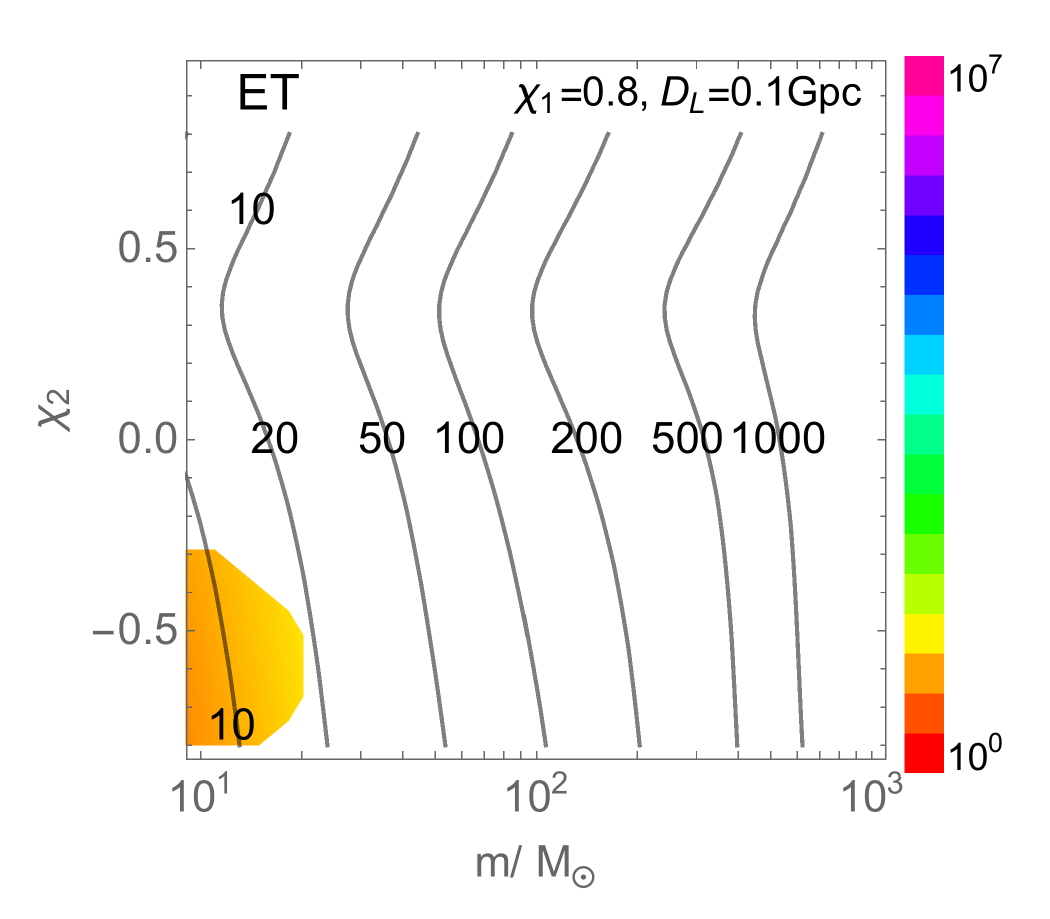} \\
\includegraphics[width=4.3cm,clip=true]{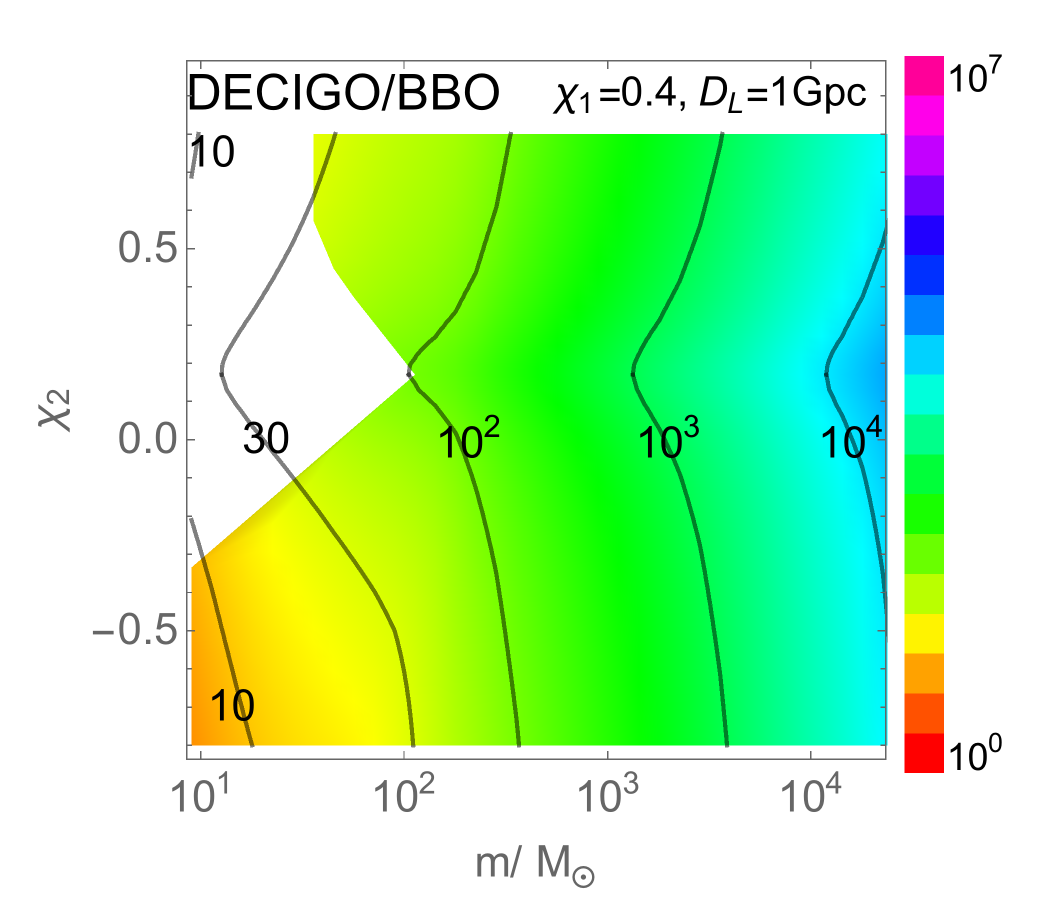}  
\includegraphics[width=4.3cm,clip=true]{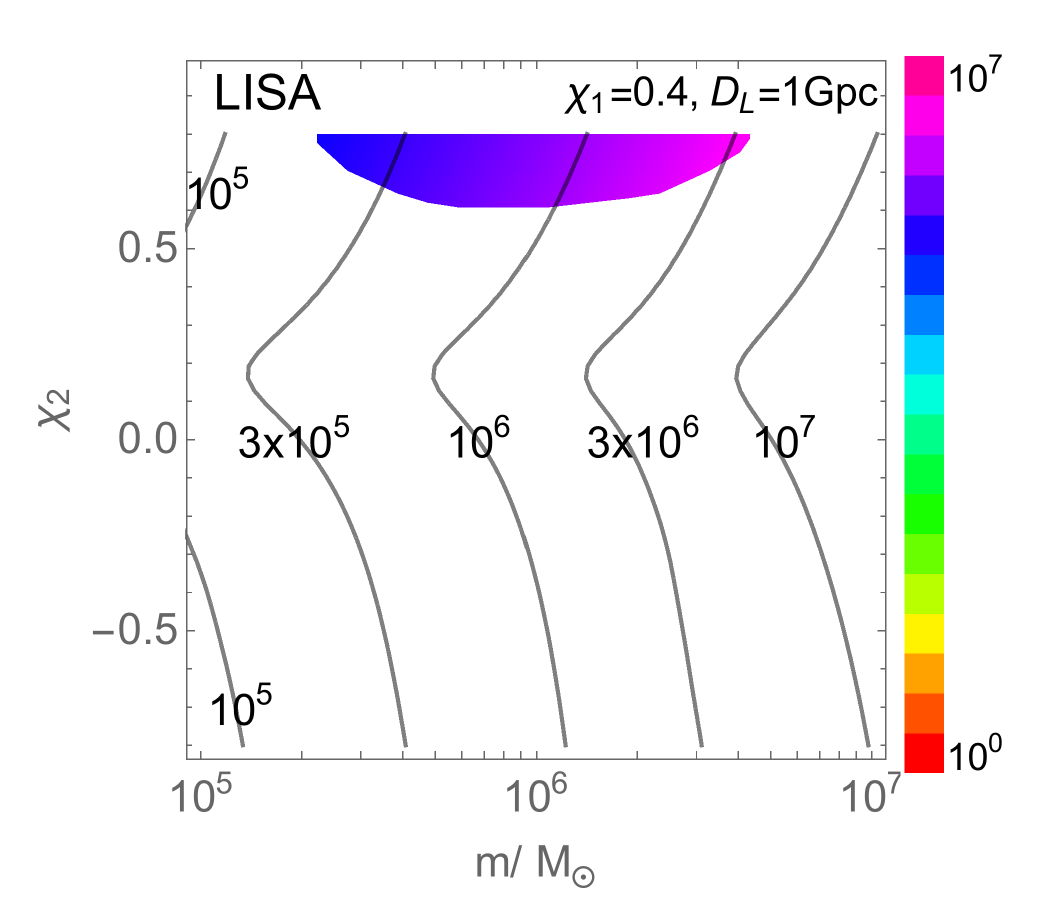} 
\end{tabular}
\caption{\label{fig2} 
Projected $1\sigma$-constraints on $\xi^{1/4}$ in km with the second-generation ground-based detectors (top left), ET (top right), DECIGO/BBO (bottom left) and LISA (bottom right) for BH binaries with $m_1/m_2=2$. The fixed values for $\chi_1$ and $D_L$ are shown at the top right of each panel. The constraints using  the second-generation ground-based detectors have been obtained by assuming that the spins are known \emph{a priori}. This roughly models projected constraints with precessing BH binary observations and may be correct within an error of 50$\%$. For other detectors, binaries are assumed to be spin-aligned (or anti-aligned). The colored contours show the regions of parameter space where the constraints on $\xi^{1/4}$ also satisfy $\Delta \zeta'<1$, and thus, the small coupling approximation is satisfied at the fiducial luminosity distances chosen.}
\end{center}
\end{figure}

To simplify the problem, we will assume that when spins precess, $\chi_{1,2}$ become uncorrelated with the other template parameters, which will suffice for an order of magnitude estimate. This amounts to performing a calculation similar to that of the spin-aligned case but assuming the spin parameters are known {\emph{a priori}}. This approximation has been shown to be very accurate, for example when considering GW bounds on the graviton Compton wavelength~\cite{stavridis}.  As an example, in the top left panel of Fig.~\ref{fig2}, we show the projected bounds on $\xi^{1/4}$ in km using second-generation ground-based detectors, where we set $(m_2/m_1,\chi_{1})=(0.5,0.8)$, set the luminosity distance to $D_L=0.1$Gpc, and also set $(\hat{\bm{S}}_1 \cdot \hat{\bm{L}}, \hat{\bm{S}}_2 \cdot \hat{\bm{L}}, \hat{\bm{S}}_1 \cdot \hat{\bm{S}}_2)=(0.5,0.5,-0.5)$. With these parameters, the CS correction to the GW phase in Eq.~\eqref{phase} vanishes at $\chi_2 \approx 0.066$, and hence one would not be able to constrain the theory around this $\chi_2$ value. The colored contours show constraints that satisfy $\Delta \zeta' < 1$ [where the smallest length scale of the system is taken to be the horizon of the smaller BH, i.e. $M=m_2( 1+\sqrt{1-\chi_2^2} )$], as otherwise the small-coupling approximation is violated. Of course, the regions that satisfy $\Delta \zeta' < 1$ depend on the choice of $D_L$: the colored contours would be larger if one detects a GW from a closer binary and thus with a higher SNR. The second-generation ground-based detectors could constrain dynamical CS gravity to  
\be
\xi^{1/4} \lesssim \mathcal{O}(10-100) \mrm{km}\,.
\ee
This is six to seven orders of magnitude stronger than current Solar System bounds~\cite{alihaimoud-chen}.

Of course, the above results are only order-of-magnitude estimates because the spin parameters do not completely decouple from other binary parameters for precessing systems, but we can quantitatively estimate its validity by following~\cite{Yagi:2009zm}. We concentrate on {\emph{simple-precessing}} systems~\cite{Apostolatos:1994mx}, where $\bm{S}_{2} = 0$ and $\hat{\bm{S}}_1 \cdot \hat{\bm{L}}$ stays constant. Since the spin-spin interaction vanishes, it is sufficient to solve the \emph{GR} precession equations to prescribe the temporal evolution of $\hat{\bm{S}}_1$ and $\hat{\bm{L}}$. We randomly generate 100 binaries over the sky, calculate the constraint from each binary and take the average. For a BH binary with $(m_1,m_2)=(5,10)M_\odot$, $\chi_1=0.8$, $\hat{\bm{S}}_1 \cdot \hat{\bm{L}}=0.5$ and $D_L=0.1$Gpc, we found $\zeta' < 61.6$ and $\xi^{1/4}<41.4$km. On the other hand, if we assume that the spins are known \emph{a priori}, these constraints become $\zeta' < 11.4$ and $\xi^{1/4}<27.1$km. Therefore, the top left panel of Fig.~\ref{fig2} may overestimate the constraints on $\zeta'$ by a factor of 5 and the colored region should shrink if we properly take the spin precession into account. However, we emphasize again that the area of this region depends on the SNR. The constraint on $\xi^{1/4}$ is less affected by our assumptions because it scales as $(\Delta \zeta')^{1/4}$, and thus, the contours in Fig.~\ref{fig2} should be accurate within an error of roughly 50$\%$. The differences tend to become smaller for larger mass parameters. Nonetheless, it would be desirable to confirm these results by first performing a more detailed Fisher analysis, taking precession properly into account and carrying out a Monte-Carlo on the sky positions, as well as performing a Bayesian model selection study. 

Next, we consider future detectors such as ET, LISA and DECIGO/BBO. For these detectors, it would be possible to constrain the theory for spin-aligned binaries within the small-coupling approximation. The results are shown in Fig.~\ref{fig2} (top right: ET, bottom left: DECIGO/BBO, bottom right: LISA).  ET and DECIGO/BBO should allow us to constrain $\xi^{1/4} \lesssim \mathcal{O}(10-100) \mrm{km}$. As expected, the constraint scales as the smallest length scale of the target system, and hence it is of $\mathcal{O}(M)$. Due to this reason, LISA can only place $\xi^{1/4} \lesssim \mathcal{O}(10^5-10^6) \mrm{km}$.

{\emph{Future Work}}.--- This paper opens the door to several follow-ups. One possibility is to include eccentricity and precessing spins in the inspiral evolutions. One could also carry out a Bayesian parameter estimation and a model-selection study to estimate projected constraints for signals with low SNR~\cite{cornishsampson}.  One could improve the waveform that is valid all the way up to merger by developing an effective-one-body resummation of the waveform.  

One could also investigate highly spinning BHs and other neutron star binaries. We have succeeded in obtaining the dipole charge for arbitrarily rapidly rotating BHs, but the waveform also depends on the deviation of the BH metric's quadrupole moment. A study of binary neutron stars in dynamical CS gravity might allow immediate constraints through binary pulsar observations. 
 
{\emph{Acknowledgments}.--- We thank Laura Sampson for checking some of the calculations using a Bayesian code. We also thank Leo Stein for useful discussions in deriving the correct expression for the dipole-dipole interaction energy. NY acknowledges support from NSF grant PHY-1114374 and NASA grant NNX11AI49G. TT is supported by the Grant-in-Aid for Scientific Research (Nos. 21244033, 21111006, 24111709, 24103006 and 24103001). 

\bibliography{master}
\end{document}